\documentclass[reprint,showpacs,prl,amsmath,amssymb,floatfix,]{revtex4-1}

\usepackage{graphicx}
\usepackage{dcolumn}
\usepackage{bm}
\usepackage{natbib}
\usepackage{color}
\usepackage{hyperref}
\begin{document}

\preprint{APS/123-QED}

\title{The root cause of hydrogen induced changes in optical transmission of vanadium}

\author{S. A. Droulias}
\author{O. Gr\aa n\"as}
\author{O. Hartmann}
\author{K. Komander}
\author{B. Hj\"orvarsson}
\author{M. Wolff}
\author{G. K. P\'alsson}
\email{Corresponding author: gunnar.palsson@physics.uu.se}
\affiliation{Division of Materials Physics and Materials Theory, Department of Physics and Astronomy, Uppsala University, Box 516, SE-75121, Uppsala Sweden}

\date{\today}

\begin{abstract}
The changes in the optical transmission of thin vanadium layers upon hydrogen absorption are found to be dominated by the volume changes of the layers and not directly linked to concentration. This effect is demonstrated by utilising the difference in the hydrogen induced expansion of V layers in Fe/V and Cr/V superlattices.  Hydrogen resides solely in the vanadium layers in these superlattices, while occupying different sites, causing different lattice expansion.  Quantitative agreement is obtained between the experimental results and first principle density functional calculations. 


\end{abstract}

\maketitle
Knowledge of the pressure-composition-isotherms of metal hydrides is crucial when studying for example hydrogen storage in metals, hydrogen embrittlement, and thin film hydrogen sensing applications~\cite{Huiberts:1996gp,Xin:2014gd,Boelsma:2017ce}. Knowledge of the concentration and its distribution in a film can also be used to measure dynamic properties of hydrogen~\cite{denBroeder:1998hqa,Palsson:2012kwa} and learn how these can change in the presence of finite-size~\cite{Huang:2017de}. In the case of thin films, optical transmission has been used to measure concentration for about two decades~\cite{Huiberts:1996gp,denBroeder:1998hqa,Borgschulte:2005hd,Dam:2007ip,Palsson:2012kwa,9928980,prinz:251910,Xin:2014gd}. The method was originally applied to systems where strong metal-insulator transitions were present, thereby simplifying the interpretation, whereby the relative volume fraction of the transparent phase is proportional to the concentration~\cite{Borgschulte:2005hd}.  In recent years, optical transmission has also been used for thin metal hydrides that do not undergo metal insulator transitions~\cite{9928980,prinz:251910}, under the assumption that the linear scaling of transmission with respect to concentration is a universal behaviour. In this work we revise the fundamental understanding of the underlying mechanism for this technique. We show that the optical transmission does not generally scale linearly with concentration, and we elaborate when this assumption hold true. Understanding the underlying mechanism for this technique is important since thermodynamic analysis rests on the assumption that optical transmission is linear with concentration.

To measure the thermodynamic properties of a hydrogen absorbing film, the optical transmission is equated to the concentration via the Lambert-Beer law~\cite{9928980,prinz:251910}.

\begin{figure}[h!]
\centering
\includegraphics[width=8.6 cm]{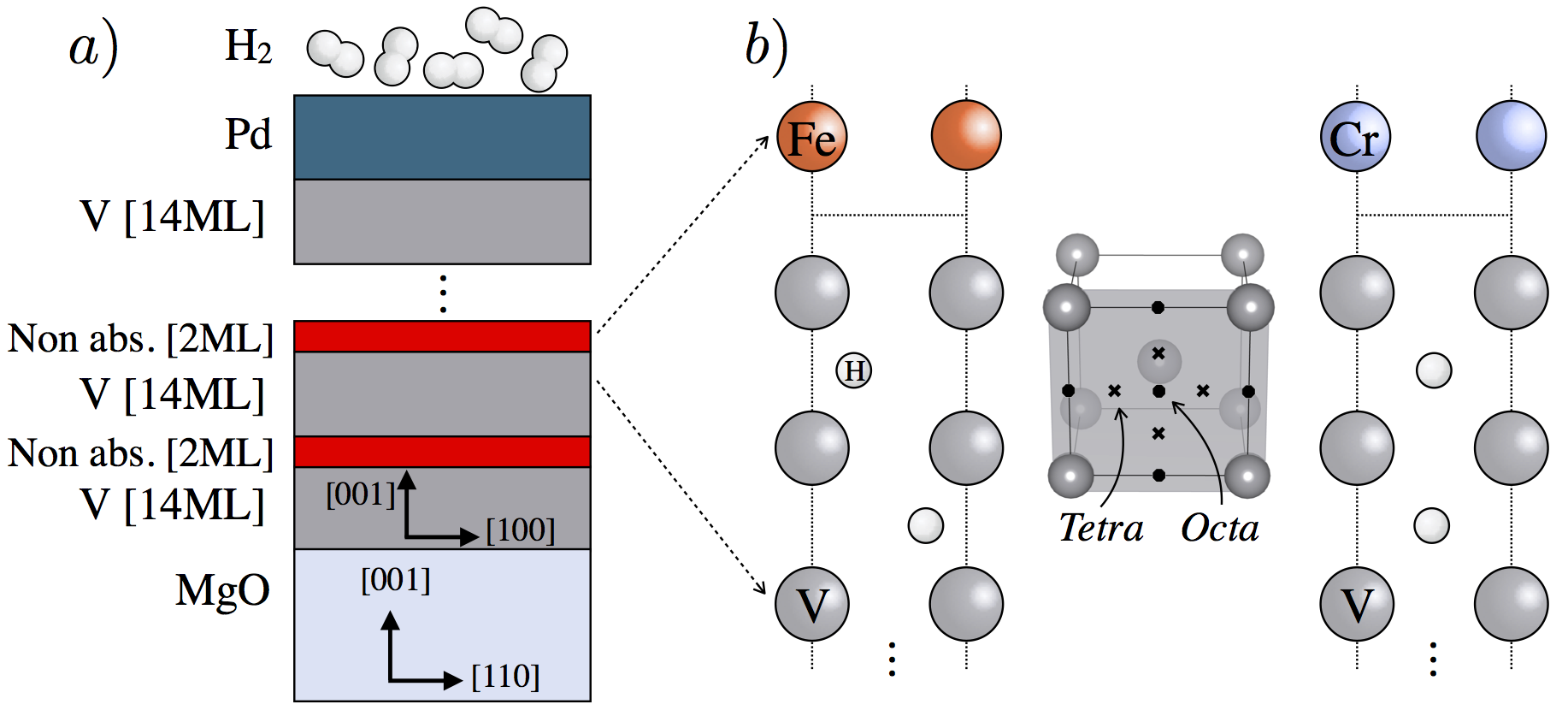}
\caption{Panel a) depicts a schematic of the superlattices including the layer thickness and their relative orientation. Panel b) shows the possible sites for hydrogen to occupy inside the two superlattices, tetrahedral or octahedral.}
\label{schematic_sl}
\end{figure}

\begin{equation}
I(c)=I_0e^{-\alpha(c,\lambda) d}\rightarrow c=(d\cdot\Delta\alpha)^{-1}\ln\frac{I(c)}{I(0)}
\label{lambert}
\end{equation}
where $I_0$ is the incident intensity, $I(c)$ is the measured transmitted intensity and hydrogen concentration $c$, $d$ is the thickness of the sample and $\alpha$ is the wavelength and concentration dependent linear absorption coefficient.
The assumptions underlying this law are that the changes in the absorption coefficient are linear with the hydrogen content and multiple reflections and scattering can be neglected.
\begin{figure*}[ht!]
\centering
\includegraphics[width=17.3 cm]{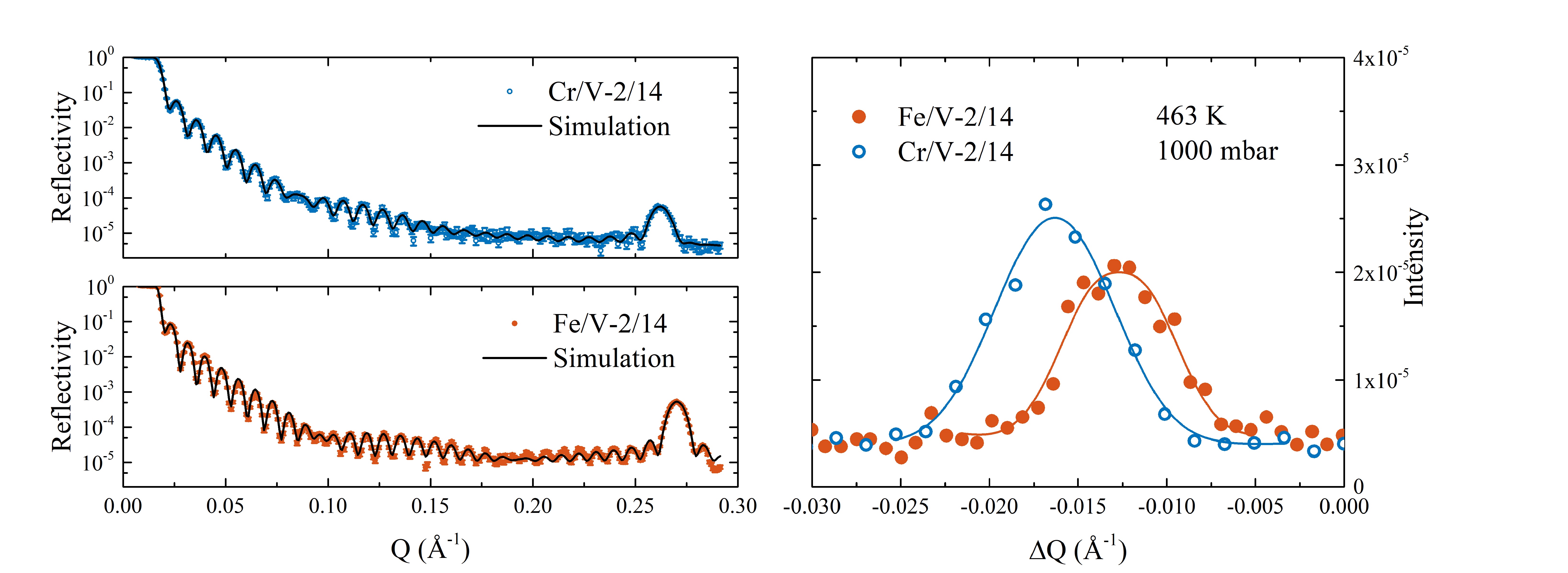}
\caption{Neutron reflectivity data from both superlattices at 463 K and 1000 mbar of D$_2$ pressure. The right panel shows the difference in position of the first superlattice satellite at the highest pressure measured.}
\label{1000mbar}
\end{figure*}
To test whether optical transmission is linear with concentration we need samples that have high crystal quality such that we can neglect effects arising from grain boundaries and other imperfections. We also would need to decouple concentration from the volume expansion, that is normally associated with uptake of hydrogen. In the present paper we therefore measure the concentration and the volume changes in-situ using neutron reflectivity, and simultaneously the optical transmission, for two samples that are identical in every respect except their expansion coefficient, $k$,
\begin{equation}
k=\frac{1}{c}\frac{\Delta V}{V}.
\end{equation}  
In Fig. 1 we show schematically a superlattice composed of a non-absorbing layer, either iron or chromium, Fe(2)/V(14) and Cr(2)/V(14) [Fe(i)/V(j) refers to $i$ atomic layers of iron and $j$ atomic layers of vanadium] with a specially tuned ratio, which corresponds to optimal crystal quality.  To arrive at this ratio, we previously investigated the influence of the ratio on the quality  and found that a ratio of 1/7 results in a fully coherent, fully strained superlattice for both Cr/V and Fe/V~\cite{Liebig:2008bx,Droulias2017}. Hence we are able to measure all the relevant quantities in-situ, under identical conditions, to determine if optical transmission scales with volume change or with the concentration. Both superlattices should have identical strain states, to avoid any changes due to difference in strain~\cite{Olsson:2005ii}. We have therefore measured the strain states in both superlattices using x-ray diffraction and applied the analysis of Birch and co-workers~\cite{Birch1995a}. We also calculate from first-principles the dielectric tensor and calculate the optical transmission using a full dynamic approach based on Parratt's algorithm to gain understanding at the level of the band structure.

Two types of superlattices were grown epitaxially using DC magnetron sputtering on 20x20 mm single-crystalline magnesium oxide substrates: MgO/[V 14 ML/Cr 2 ML]$_{25}$/V 14 ML/Pd 7 nm and MgO/[V 14 ML/Fe 2 ML]$_{30}$/V 14 ML/Pd 7 nm where monolayer is abbreviated as ML. The growth and detailed characterisation of the Fe/V samples, including Fe(2)/V(14), can be found in Ref.~\onlinecite{Droulias2017} but are summarised as follows. The samples were grown at a base pressure of 3$\times$10$^{-7}$ Pa and an argon purity of 99.9997$\%$. The purity of all targets was 99.99$\%$. The structural characterisation was done using x-ray reflectivity and diffraction where both superlattices exhibited high crystal quality and were found to have mosaic spread of 0.03$^{\textrm{o}}$ and an out-of-plane coherence length equal to the total thickness, and the in-plane coherence length was found to be several hundred nano meters. We have found previously that analyzing the position and intensity of the Bragg peaks only, leads to misleading results, and hence both neutron and x-ray reflectivity measurements were fitted using GenX \cite{Bjorck:aj5091}, which confirmed excellent layering and flatness of the samples at mesoscopic length scales.\\

The neutron measurements were done at the Institut Laue Langevin, Grenoble, France, using the reflectometer SuperADAM, with a wavelength of 0.5183 nm \cite{superadam}.  An ultra high vacuum chamber, with a base pressure of less than $5\times10^{-9}$ mbar was used to minimise the influence of impurities on the experiment. The deuterium pressure was measured using capacitance gauges. The chamber is made of aluminium with explosion welded stainless steel flanges, to allow for ultra high vacuum conditions using standard Cu gaskets. Part of the chamber is thinned to optimise neutron transmission and includes optical windows made of sapphire for simultaneous light transmission. The thermodynamic properties were measured using optical transmission using the equipment described in~\cite{prinz:251910} with a wavelength of $\lambda=625$ nm. All measurements were made using ultra pure hydrogen, purified by a Nu-pure purifier, whereby ppb impurties can be achived. Heating was done externally using a customised heating jacket from Hemi heating to minimise any thermal gradients inside the chamber. The temperature of the sample was monitored using a thermocouple feed-through in contact with the backside of the substrate. Before the neutron reflectivity curve was measured, and after changing the pressure, enough time was given for equilibrium to settle in. Equilibrium was monitored by observing the pressure, temperature and the optical transmission.

The deuterium concentration,  $c=N_\mathrm{D}/N_\mathrm{V}$ where $N_{\mathrm{V,(D)}}$ is the number of vanadium (deuterium) atoms in the volume of a unit cell of the superlattice, can be obtained from the expression for the scattering length density of the deuterium containing vanadium layer:
\begin{equation}
\rho_{\mathrm{VD_c}}(c) = \frac{\textsc{N}_\textrm{V} \textrm{b}_\textrm{V}}{V(c)} + \frac{\textsc{N}_\textrm{D}(c) \textrm{b}_\textrm{D}}{V(c)}
\label{rho}
\end{equation}
where $ \rho_{VD_c} $ is the scattering length density of the absorbing layer, $\textrm{N}_\textrm{V(D)}$ is the number of atoms of vanadium or deuterium within the concentration dependent volume $V(c)$, and $\textrm{b}_\textrm{V(D)}$ are the bound coherent neutron scattering lengths. The superlattices are clamped by the substrate, which implies that the volume expansion due to hydrogenation is equal to the expansion of the vanadium layer in the out-of-plane direction. The concentration can be calculated in a straight forward manner from eq.~\ref{rho} and with $\rho_{\mathrm{VD_0}}(0)=\rho_\mathrm{V}$ and is:

\begin{equation}
\textrm{c} = \left[\frac{\rho_{\mathrm{VD_c}}(c)}{\rho_{\mathrm{V}}}\left(1+\frac{V(c)-V(0)}{V(0)}\right)-1\right]\frac{\textrm{b}_\mathrm{V}}{\textrm{b}_\mathrm{D}}
\end{equation}
The concentration is found by using the tabulated values for $\textrm{b}_\textrm{V(D)}$ whereas the scattering length densities and the volume expansion are obtained from fits of the neutron reflectivity curve at each pressure.

Using the analysis presented in Birch et. al. \cite{Birch1995a} the strain state of the V layers was investigated in the two superlattices, Cr(2)/V(14) and Fe(2)/V(14). Fully coherent V layers, grown epitaxially on MgO, will have an in-plane lattice parameter of $4.211/\sqrt{2}$=2.98 \AA. This registry with the substrate will hold for both elements comprising the superlattice and for both cases of Cr/V and Fe/V. The out of plane lattice parameters of the individual layers will depend on the elastic constants of the individual materials and we find the strain state of the V layer to be identical in both cases within  2$\%$. For a more extensive description of the calculation of strain we refer to [\onlinecite{Droulias2017}].  We have thus demonstrated that the average strain state in the vanadium layers is identical within the uncertainty in these two systems.

The left panel of Fig.~\ref{1000mbar} shows neutron reflectivity of superlattices, for Cr(2)/V(14) and Fe(2)/V(14) without deuterium. The modulation in intensity, seen throughout the whole range of the scans, are Kiessig fringes and originate from constructive and destructive interference of the neutron from the substrate interface and the surface. The first satellite of the superlattice is at $Q\approx0.26$ \AA$^{-1}$, from which the chemical repeat distance can be obtained. The neutron data confirms the excellent structural quality of the samples.

The samples were exposed to a deuterium atmosphere of varying pressures and for each pressure, a reflectivity pattern was measured. Fig~\ref{1000mbar} shows the change in the position of the first satellite for the two superlattices at 1000 mbar. As can be seen the peaks have shifted (deuterium-induced expansion) by different amounts and exhibit different intensities, indicating a difference in the deuterium absorption behaviour. The intensity of the peak depends primarily on the difference in scattering length density between the layers comprising the superlattice. Due to the scattering length density of the deuterated layer changing with deuterium content, the contrast between the layers is also altered. Therefore, the concentration of deuterium in the vanadium layers can be determined by using the intensity of the peak to calculate the scattering length density of the absorbing layer. In this work, the fitting tool GenX~\cite{Bjorck:aj5091} was used to fit the whole reflectivity pattern of all pressures.
\begin{figure}
\includegraphics[width=7.0 cm]{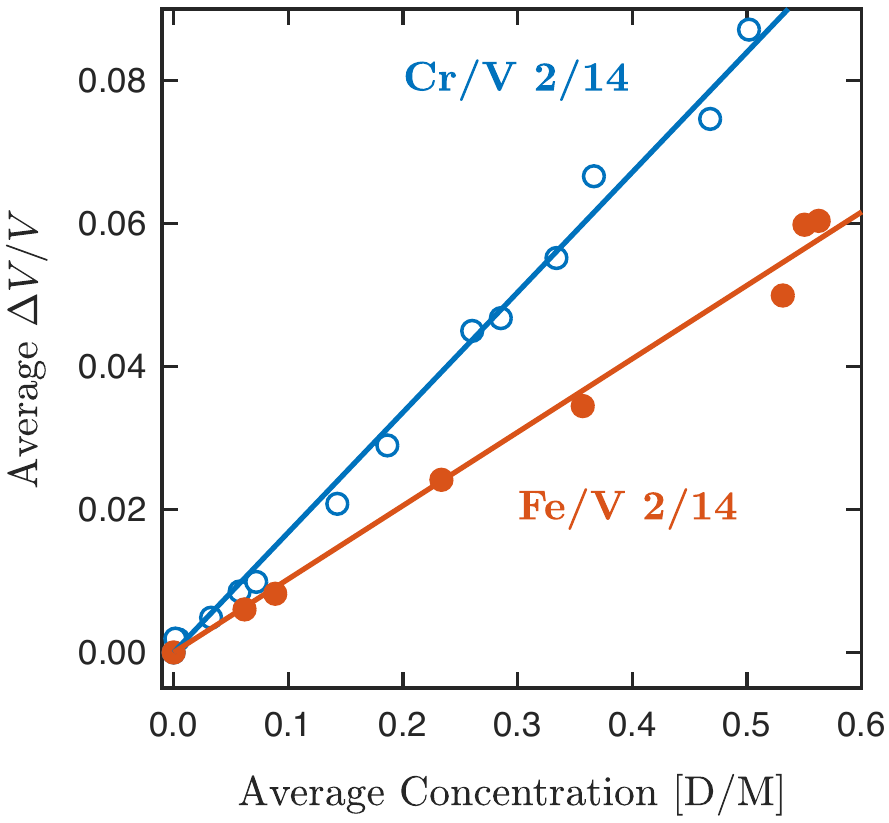}
\caption{The figure shows the scaling between deuterium concentration and volume changes. The difference in slope is evidence of different site occupancy.}
\label{cvsv}
\end{figure}

\begin{figure}
\includegraphics[width=7.0 cm]{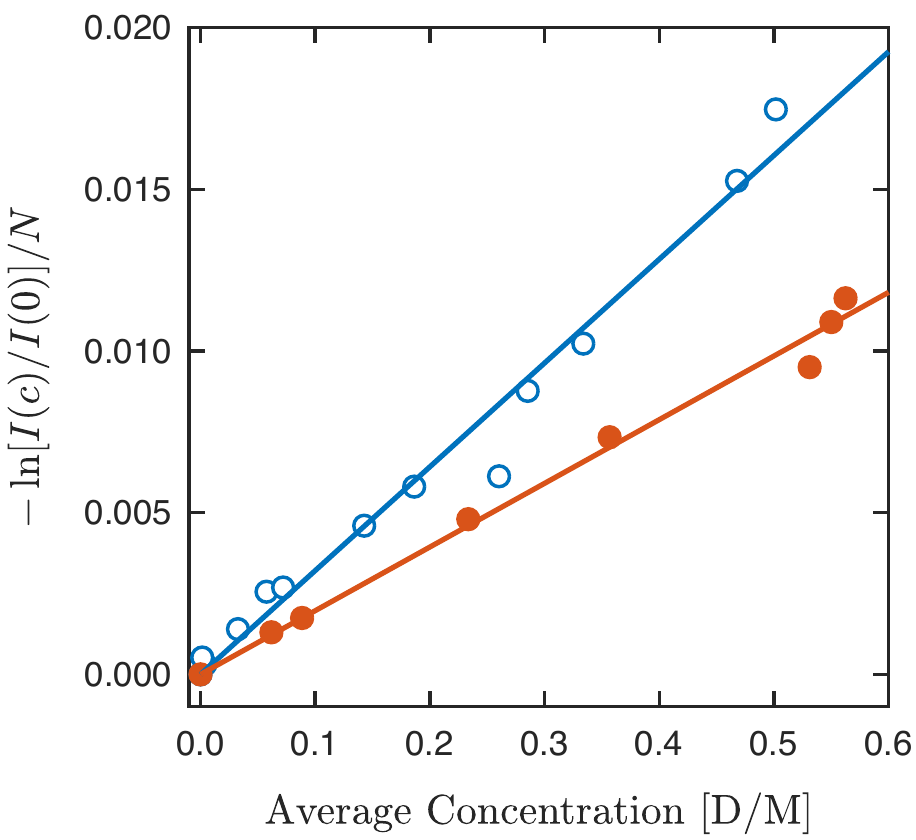}
\caption{The figure shows the relationship between deuterium concentration and optical transmission, which is linear for both Fe/V 2/14 and Cr/V 2/14 however the slopes are quite different. The optical transmission has been scaled by the number of repeats in each superlattice to compensate for a slight difference in thickness.}
\label{cvsT}
\end{figure}

\begin{figure}
\includegraphics[width=6.0 cm]{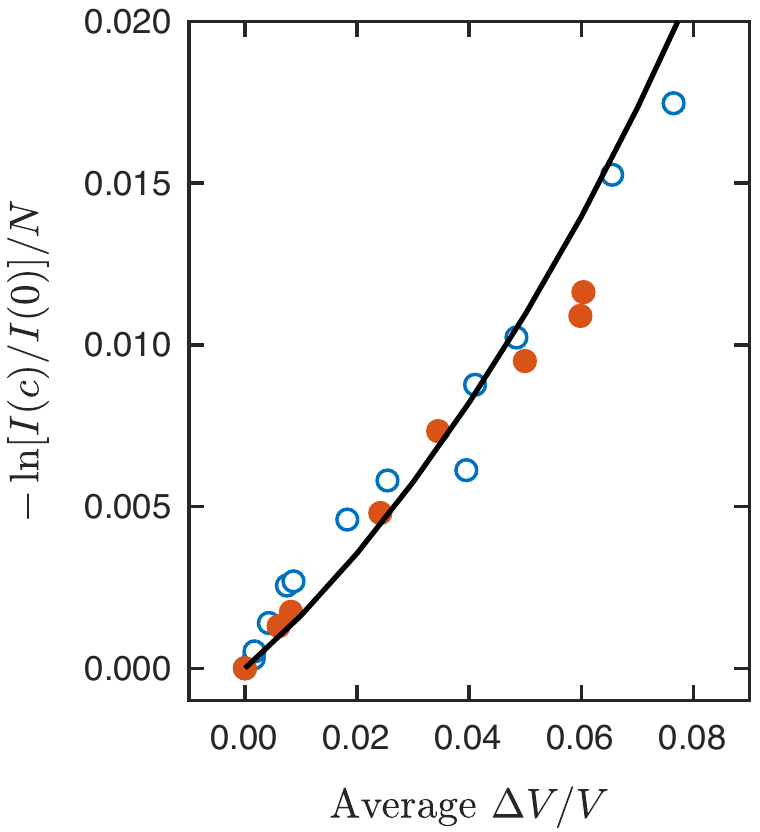}
\caption{A universal curve emerges between volume change is compared with optical transmission. The solid line is the result from the density functional calculations and excellent agreement is found between theory and experiment. }
\label{vvsT}
\end{figure}

\begin{figure}[h!]
\centering
\includegraphics[width=8.76 cm]{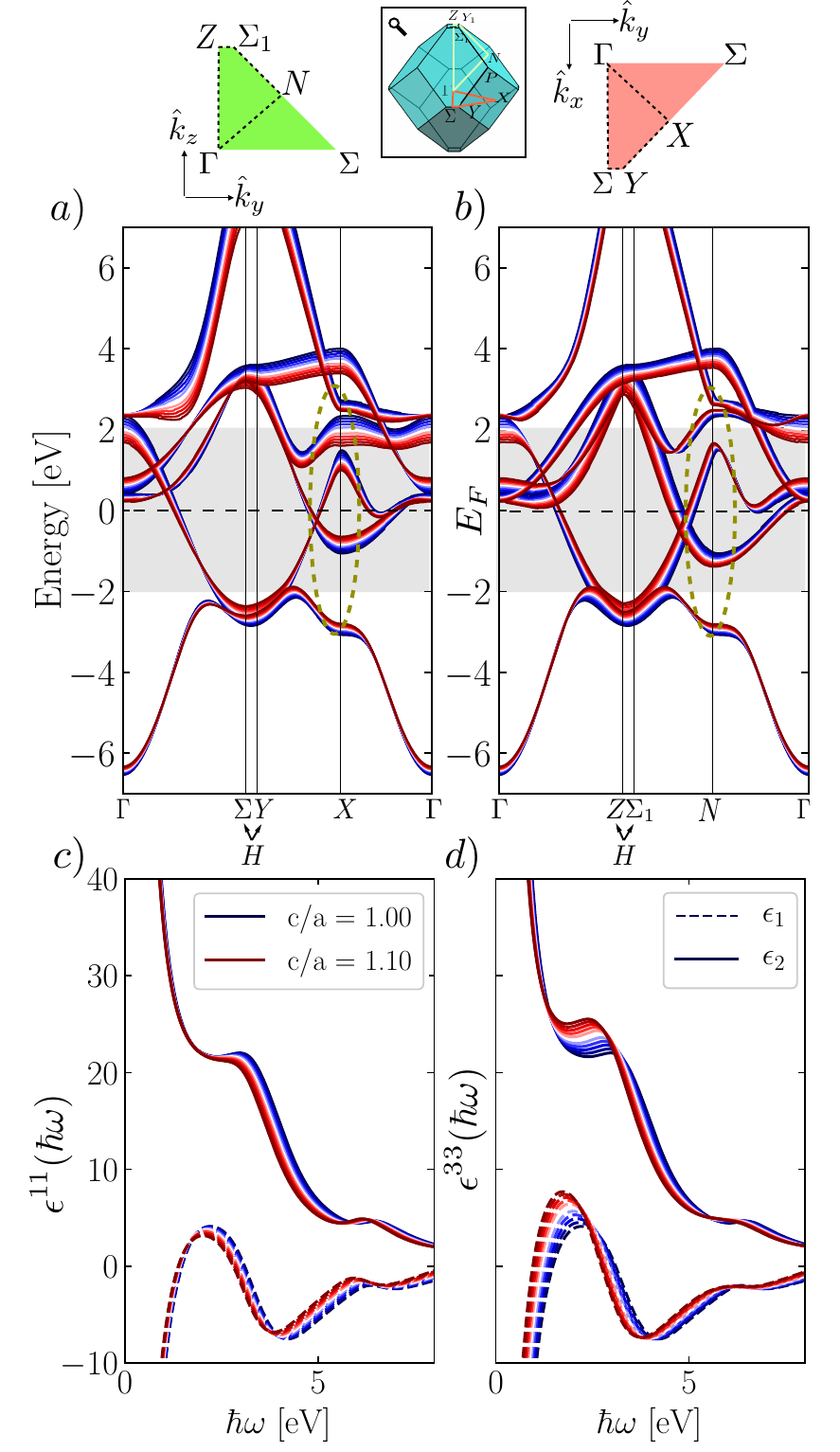}
\caption{The upper panel shows the influence of changing the c/a ratio on the band structure as obtained from the all-electron calculation. The bottom panels show the real and imaginary parts of the dielectric tensor for both in-plane and out-of plane components as a function of the c/a ratio.}
\label{elkdielectric}
\end{figure}
Figure ~\ref{cvsv} shows the deuterium concentration versus the volume change of the two superlattices. Markedly different slopes are seen, which confirms the fact that the two superlattices have different expansion coefficients $k$, most likely due to difference in site occupancy. We now have two samples that are very similar in every respect expect for the specific volume change and can now go ahead and compare how the optical transmission scales with concentration. Figure  \ref{cvsT} shows the deuterium concentration versus the simultaneously determined optical transmission for the two superlattices. As can be seen in the Fig. the transmitted intensity is indeed linear for both superlattices. The optical transmission has been scaled by the number of periods in each superlattice to account for a slight change in total thickness. Note that the slope is different between the two superlattices, which shows that occupancy influences the conversion factor between optical transmission and concentration. We see that the there is a strong correlation between volume change and optical transmission, which suggests a common cause.

Figure~\ref{vvsT} shows the optical transmission plotted against the volume change and we can immediately see that a universal behaviour emerges. When hydrogen enters the crystal structure and occupies an interstitial site, the subsequent s-d hybridization between the hydrogen 1s electron and the vanadium 3d electron, weakens the metal-metal bonds and leads to an expansion of the structure. Optical properties are in principle sensitive to both the rearrangement of the electronic states caused by hydrogen as well as the change in electron density accompanying the expansion. 

To further understand the influence of volume expansion on optical transmission we have undertaken a first-principle study to calculate the dielectric tensor as a function of c/a ratio. 
We calculate the dielectric function for a single unit cell of vanadium for $c/a$ ratios going from 1.0 to 1.1 while keeping the in-plane lattice parameters fixed at the experimentally clamped values. This mimics both the one dimensional nature of the volume change in the superlattices as well as the expansion caused by hydrogen. In this way we investigate only the volume aspects of the hydrogen absorption.  To this end, we use the full-potential all-electron electronic structure code Elk \cite{elk}. Effects of linear optics from ground-state Kohn-Sham calculations for the full dielectric tensor is made, as well as Time-dependent density functional theory (TD-DFT) within linear-response and with full local fields corrections for trace dielectric properties, $\epsilon_{1}$, $\epsilon_{2}$. Convergence with respect to $k$-point sampling, $G$-vector cut-off, basis set and potential expansion was made. The underlying DFT calculations are converged at $R_{\mathrm{min}}^{\mathrm{MT}}\times \mathrm{max}\{|G+k|\}=8$, maximum $l$-quantum number for the APW expansion at 10, maximum $l$-quantum number for expanding density and potential within the muffin-tin at 10, density and potential $G$-vector cutoff at 12.0 $a.u.^{-1}$, and $24^{3}$ k-points. For the trace of the dielectric tensor, TD-DFT was used including local-field effects up to a cut-off of 3 $a.u.^{-1}$ $G$-vectors. The full dielectric tensor is calculated with linear optics within the random phase approximation in the $q\rightarrow 0$ limit. The two contributions to the dielectric function, intraband transitions and interband transition are both electron density dependent, the former having the expression
\begin{equation}
\epsilon^{(1)}_{\textrm{intra}}=1-\frac{\omega_p^2}{\omega^2+\gamma^2}\qquad \epsilon^{(2)}_{\textrm{intra}}=\frac{\gamma\omega_p^2}{\omega^3+\omega\gamma^2}\quad\omega_p=\sqrt{\frac{n_ee^2}{m^*\epsilon_0}}
\end{equation}
where $\omega_p$ is the free electron plasma frequency, $n_e$ is the electron density, $m^*$ is the effective mass, $\gamma$ is the relaxation time and $\epsilon_0$ is the permittivity of free space and $e$ is the elementary electron charge. In the calculations we calculated the intraband term using the electron density obtained by the self consistent all electron calculation~\cite{Romaniello:2006ve} and assumed a value of $\gamma=0.5$ eV. The same smearing was used for the interband transitions. The smearing width was taken to be the same as determined by Laurent~\cite{Laurent:1978jq} which is somewhat larger than used by Weaver~\cite{Weaver:1974vx} of 0.5 eV. Sacchetti calculated the electron-electron interaction in vanadium and supported using the value of Laurent~\cite{Sacchetti:1983jl}. 

Figure \ref{fig:opticscompare} in the appendix shows the experimental bulk dielectric function~\cite{Palik:1998} as well as the results from the present calculation for zero tetragonal strain. As can be seen in Fig. \ref{fig:opticscompare} the effects of local fields is small. The agreement with the experimental data is good even though electron-phonon coupling and energy dependent smearing is not included. Our results are very similar to the results of Romaniello and co-workers~\cite{Romaniello:2006ve}, further strengthening the approach used here. From the dielectric tensor, the refractive index and the extinction coefficient were calculated and the optical transmission subsequently determined assuming a 500 \AA~ vanadium film on MgO using Parratt's dynamical formalism. The experimental values for the refractive index of MgO was used.  

The solid black line in the middle panel of Fig.~\ref{vvsT} shows the calculated optical transmission as a function of the change in volume in this clamped configuration from the theoretical calculations. We obtain excellent agreement with the experimental data, which strongly suggests that volume effects drive the observed changes in transmission when hydrogen is absorbed, rather than rearrangements of electronic states around the Fermi level. This would imply that for metals that do not undergo metal-insulator transitions, the linearity between concentration and optical transmission is due to changes in the electron density. This finding is also consistent with the fact that most of the changes to the electronic structure due to hydrogen absorption takes place around 7 eV below the Fermi level, where the spectral weight owing to s-d hybridization is located~\cite{Smithson:2002tla,PhysRevB.90.045420}. Hence for optical measurements in the visible range, such changes do not contribute to the optical response.

Fig.~\ref{elkdielectric} shows the changes in the band structure in selected symmetry directions as the vanadium crystal is stretched, while keeping the in-plane lattice parameter constant, as well as the subsequent changes to the dielectric function. It is evident from Fig. \ref{elkdielectric} that mono-crystallinity is important if optical transmission is to be used, since the changes in the out-of-plane (longitudinal) components vary substantially more than the transverse, due to changes in the band-structure. As the volume expands, and the film is strained, the orbital overlap is changed, inducing a change in band-width. The trends regarding the non-trivial changes in the optical properties, as produced by first-principles calculations, can be of guidance to determine the most suitable optical frequencies to measure structural changes. 

We find using simultaneous neutron reflectivity and optical transmission that the hydrogen concentration scales linearly with optical transmission at $\lambda=625$ nm for both superlattices but with different slopes. The origin of this linearity is traced to the changes in volume that cause changes in the electron density, which is confirmed by first principle calculations. We generalise these findings to that any (non-complex) metal hydride film, even without exhibiting metal-insulator transitions during hydrogenation can be measured using optical transmission only as long the expansion coefficient of the hydride does not change. Our conclusion would imply that any metal hydride measurements where optical transmission has been used and for instance hydrogen occupancy has changed during absorption, needs to be revisited.
\section{Acknowledgements}
The work of O.G. was supported by the Strategic Research Foundation (SSF) Grant No. ICA16-0037.  
Computational resources where provided by NSC under the allocation SNIC 2018-3-221.
\section{Appendix}
\begin{figure}
\begin{center}
\includegraphics[width=8.6cm]{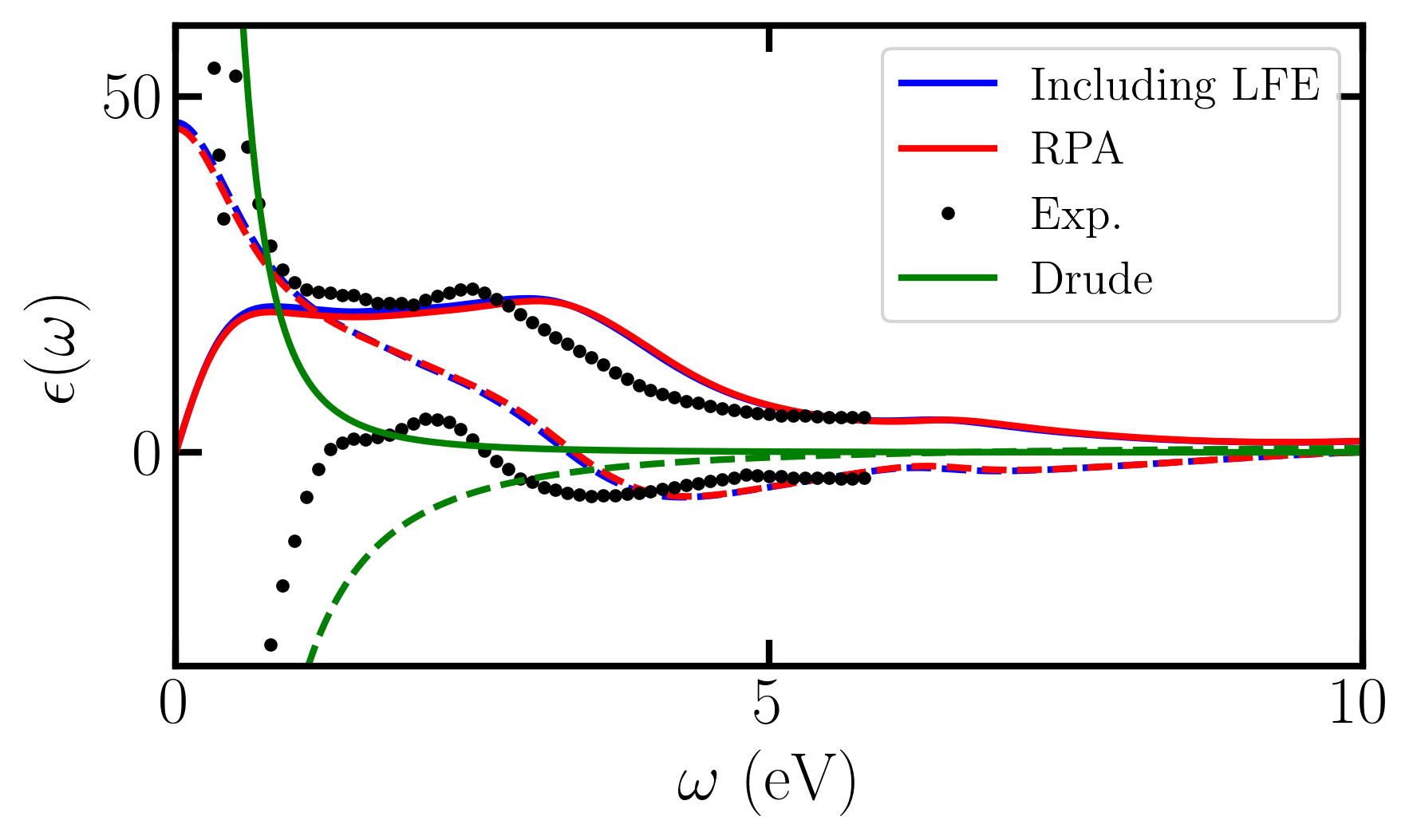}
\caption{A comparison between the trace of the dielectric tensor calculated with different level of theory, and the bulk experimental dielectric function of vanadium~\cite{Palik:1998}. $\epsilon_{1}$ is dashed and $\epsilon_{2}$  solid. The Drude contribution is plotted separately in order to clarify the different contributions. The computational results show a blue-shift of the main absorption shoulder, but all general features compare well.}
\label{fig:opticscompare}
\end{center}
\end{figure}

Figure~\ref{fig:opticscompare} shows how the calculated dielectric function compares with unstrained experimental bulk measurements at room temperature. We provide this comparison to verify that the theoretical approach is close enough to reality to be meaningful for comparing changes in optical properties with uniaxial expansion.
\bibliography{Palsson2018}

\begin{thebibliography}{24}%
\makeatletter
\providecommand \@ifxundefined [1]{%
 \@ifx{#1\undefined}
}%
\providecommand \@ifnum [1]{%
 \ifnum #1\expandafter \@firstoftwo
 \else \expandafter \@secondoftwo
 \fi
}%
\providecommand \@ifx [1]{%
 \ifx #1\expandafter \@firstoftwo
 \else \expandafter \@secondoftwo
 \fi
}%
\providecommand \natexlab [1]{#1}%
\providecommand \enquote  [1]{``#1''}%
\providecommand \bibnamefont  [1]{#1}%
\providecommand \bibfnamefont [1]{#1}%
\providecommand \citenamefont [1]{#1}%
\providecommand \href@noop [0]{\@secondoftwo}%
\providecommand \href [0]{\begingroup \@sanitize@url \@href}%
\providecommand \@href[1]{\@@startlink{#1}\@@href}%
\providecommand \@@href[1]{\endgroup#1\@@endlink}%
\providecommand \@sanitize@url [0]{\catcode `\\12\catcode `\$12\catcode
  `\&12\catcode `\#12\catcode `\^12\catcode `\_12\catcode `\%12\relax}%
\providecommand \@@startlink[1]{}%
\providecommand \@@endlink[0]{}%
\providecommand \url  [0]{\begingroup\@sanitize@url \@url }%
\providecommand \@url [1]{\endgroup\@href {#1}{\urlprefix }}%
\providecommand \urlprefix  [0]{URL }%
\providecommand \Eprint [0]{\href }%
\providecommand \doibase [0]{http://dx.doi.org/}%
\providecommand \selectlanguage [0]{\@gobble}%
\providecommand \bibinfo  [0]{\@secondoftwo}%
\providecommand \bibfield  [0]{\@secondoftwo}%
\providecommand \translation [1]{[#1]}%
\providecommand \BibitemOpen [0]{}%
\providecommand \bibitemStop [0]{}%
\providecommand \bibitemNoStop [0]{.\EOS\space}%
\providecommand \EOS [0]{\spacefactor3000\relax}%
\providecommand \BibitemShut  [1]{\csname bibitem#1\endcsname}%
\let\auto@bib@innerbib\@empty
\bibitem [{\citenamefont {Huiberts}\ \emph {et~al.}(1996)\citenamefont
  {Huiberts}, \citenamefont {Griessen}, \citenamefont {Rector}, \citenamefont
  {Wijngaarden}, \citenamefont {Dekker}, \citenamefont {de~Groot},\ and\
  \citenamefont {Koeman}}]{Huiberts:1996gp}%
  \BibitemOpen
  \bibfield  {author} {\bibinfo {author} {\bibfnamefont {J.~N.}\ \bibnamefont
  {Huiberts}}, \bibinfo {author} {\bibfnamefont {R.}~\bibnamefont {Griessen}},
  \bibinfo {author} {\bibfnamefont {J.~H.}\ \bibnamefont {Rector}}, \bibinfo
  {author} {\bibfnamefont {R.~J.}\ \bibnamefont {Wijngaarden}}, \bibinfo
  {author} {\bibfnamefont {J.~P.}\ \bibnamefont {Dekker}}, \bibinfo {author}
  {\bibfnamefont {D.~G.}\ \bibnamefont {de~Groot}}, \ and\ \bibinfo {author}
  {\bibfnamefont {N.~J.}\ \bibnamefont {Koeman}},\ }\href@noop {} {\bibfield
  {journal} {\bibinfo  {journal} {Nature}\ }\textbf {\bibinfo {volume} {380}},\
  \bibinfo {pages} {231} (\bibinfo {year} {1996})}\BibitemShut {NoStop}%
\bibitem [{\citenamefont {Xin}\ \emph {et~al.}(2014)\citenamefont {Xin},
  \citenamefont {Palsson},\ and\ \citenamefont {Hjorvarsson}}]{Xin:2014gd}%
  \BibitemOpen
  \bibfield  {author} {\bibinfo {author} {\bibfnamefont {X.}~\bibnamefont
  {Xin}}, \bibinfo {author} {\bibfnamefont {G.~K.}\ \bibnamefont {Palsson}}, \
  and\ \bibinfo {author} {\bibfnamefont {B.}~\bibnamefont {Hjorvarsson}},\
  }\href@noop {} {\bibfield  {journal} {\bibinfo  {journal} {Physical Review
  Letters}\ }\textbf {\bibinfo {volume} {113}},\ \bibinfo {pages} {046103}
  (\bibinfo {year} {2014})}\BibitemShut {NoStop}%
\bibitem [{\citenamefont {Boelsma}\ \emph {et~al.}(2017)\citenamefont
  {Boelsma}, \citenamefont {Bannenberg}, \citenamefont {Van~Setten},
  \citenamefont {Steinke}, \citenamefont {van Well},\ and\ \citenamefont
  {Dam}}]{Boelsma:2017ce}%
  \BibitemOpen
  \bibfield  {author} {\bibinfo {author} {\bibfnamefont {C.}~\bibnamefont
  {Boelsma}}, \bibinfo {author} {\bibfnamefont {L.~J.}\ \bibnamefont
  {Bannenberg}}, \bibinfo {author} {\bibfnamefont {M.~J.}\ \bibnamefont
  {Van~Setten}}, \bibinfo {author} {\bibfnamefont {N.~J.}\ \bibnamefont
  {Steinke}}, \bibinfo {author} {\bibfnamefont {A.~A.}\ \bibnamefont {van
  Well}}, \ and\ \bibinfo {author} {\bibfnamefont {B.}~\bibnamefont {Dam}},\
  }\href@noop {} {\bibfield  {journal} {\bibinfo  {journal} {Nature
  Communications}\ }\textbf {\bibinfo {volume} {8}},\ \bibinfo {pages} {15718}
  (\bibinfo {year} {2017})}\BibitemShut {NoStop}%
\bibitem [{\citenamefont {den Broeder}\ \emph {et~al.}(1998)\citenamefont {den
  Broeder}, \citenamefont {van~der Molen}, \citenamefont {Kremers},
  \citenamefont {Huiberts}, \citenamefont {Nagengast}, \citenamefont {van
  Gogh}, \citenamefont {Huisman}, \citenamefont {Koeman}, \citenamefont {Dam},
  \citenamefont {Rector}, \citenamefont {Plota}, \citenamefont {Haaksma},
  \citenamefont {Hanzen}, \citenamefont {Jungblut}, \citenamefont {Duine},\
  and\ \citenamefont {Griessen}}]{denBroeder:1998hqa}%
  \BibitemOpen
  \bibfield  {author} {\bibinfo {author} {\bibfnamefont {F.~J.~A.}\
  \bibnamefont {den Broeder}}, \bibinfo {author} {\bibfnamefont {S.~J.}\
  \bibnamefont {van~der Molen}}, \bibinfo {author} {\bibfnamefont
  {M.}~\bibnamefont {Kremers}}, \bibinfo {author} {\bibfnamefont {J.~N.}\
  \bibnamefont {Huiberts}}, \bibinfo {author} {\bibfnamefont {D.~G.}\
  \bibnamefont {Nagengast}}, \bibinfo {author} {\bibfnamefont {A.~T.~M.}\
  \bibnamefont {van Gogh}}, \bibinfo {author} {\bibfnamefont {W.~H.}\
  \bibnamefont {Huisman}}, \bibinfo {author} {\bibfnamefont {N.~J.}\
  \bibnamefont {Koeman}}, \bibinfo {author} {\bibfnamefont {B.}~\bibnamefont
  {Dam}}, \bibinfo {author} {\bibfnamefont {J.~H.}\ \bibnamefont {Rector}},
  \bibinfo {author} {\bibfnamefont {S.}~\bibnamefont {Plota}}, \bibinfo
  {author} {\bibfnamefont {M.}~\bibnamefont {Haaksma}}, \bibinfo {author}
  {\bibfnamefont {R.~M.~N.}\ \bibnamefont {Hanzen}}, \bibinfo {author}
  {\bibfnamefont {R.~M.}\ \bibnamefont {Jungblut}}, \bibinfo {author}
  {\bibfnamefont {P.~A.}\ \bibnamefont {Duine}}, \ and\ \bibinfo {author}
  {\bibfnamefont {R.}~\bibnamefont {Griessen}},\ }\href@noop {} {\bibfield
  {journal} {\bibinfo  {journal} {Nature}\ }\textbf {\bibinfo {volume} {394}},\
  \bibinfo {pages} {656} (\bibinfo {year} {1998})}\BibitemShut {NoStop}%
\bibitem [{\citenamefont {Palsson}\ \emph {et~al.}(2012)\citenamefont
  {Palsson}, \citenamefont {Bliersbach}, \citenamefont {Wolff}, \citenamefont
  {Zamani},\ and\ \citenamefont {Hjorvarsson}}]{Palsson:2012kwa}%
  \BibitemOpen
  \bibfield  {author} {\bibinfo {author} {\bibfnamefont {G.~K.}\ \bibnamefont
  {Palsson}}, \bibinfo {author} {\bibfnamefont {A.}~\bibnamefont {Bliersbach}},
  \bibinfo {author} {\bibfnamefont {M.}~\bibnamefont {Wolff}}, \bibinfo
  {author} {\bibfnamefont {A.}~\bibnamefont {Zamani}}, \ and\ \bibinfo {author}
  {\bibfnamefont {B.}~\bibnamefont {Hjorvarsson}},\ }\href@noop {} {\bibfield
  {journal} {\bibinfo  {journal} {Nature Communications}\ }\textbf {\bibinfo
  {volume} {3}},\ \bibinfo {pages} {892} (\bibinfo {year} {2012})}\BibitemShut
  {NoStop}%
\bibitem [{\citenamefont {Huang}\ \emph {et~al.}(2017)\citenamefont {Huang},
  \citenamefont {Palsson}, \citenamefont {Brischetto}, \citenamefont {Palonen},
  \citenamefont {Droulias}, \citenamefont {Hartmann}, \citenamefont {Wolff},\
  and\ \citenamefont {Hjorvarsson}}]{Huang:2017de}%
  \BibitemOpen
  \bibfield  {author} {\bibinfo {author} {\bibfnamefont {W.}~\bibnamefont
  {Huang}}, \bibinfo {author} {\bibfnamefont {G.~K.}\ \bibnamefont {Palsson}},
  \bibinfo {author} {\bibfnamefont {M.}~\bibnamefont {Brischetto}}, \bibinfo
  {author} {\bibfnamefont {H.}~\bibnamefont {Palonen}}, \bibinfo {author}
  {\bibfnamefont {S.~A.}\ \bibnamefont {Droulias}}, \bibinfo {author}
  {\bibfnamefont {O.}~\bibnamefont {Hartmann}}, \bibinfo {author}
  {\bibfnamefont {M.}~\bibnamefont {Wolff}}, \ and\ \bibinfo {author}
  {\bibfnamefont {B.}~\bibnamefont {Hjorvarsson}},\ }\href@noop {} {\bibfield
  {journal} {\bibinfo  {journal} {New Journal of Physics}\ }\textbf {\bibinfo
  {volume} {19}},\ \bibinfo {pages} {123004} (\bibinfo {year}
  {2017})}\BibitemShut {NoStop}%
\bibitem [{\citenamefont {Borgschulte}\ \emph {et~al.}(2005)\citenamefont
  {Borgschulte}, \citenamefont {Lohstroh}, \citenamefont {Westerwaal},
  \citenamefont {Schreuders}, \citenamefont {Rector}, \citenamefont {Dam},\
  and\ \citenamefont {Griessen}}]{Borgschulte:2005hd}%
  \BibitemOpen
  \bibfield  {author} {\bibinfo {author} {\bibfnamefont {A.}~\bibnamefont
  {Borgschulte}}, \bibinfo {author} {\bibfnamefont {W.}~\bibnamefont
  {Lohstroh}}, \bibinfo {author} {\bibfnamefont {R.}~\bibnamefont
  {Westerwaal}}, \bibinfo {author} {\bibfnamefont {H.}~\bibnamefont
  {Schreuders}}, \bibinfo {author} {\bibfnamefont {J.}~\bibnamefont {Rector}},
  \bibinfo {author} {\bibfnamefont {B.}~\bibnamefont {Dam}}, \ and\ \bibinfo
  {author} {\bibfnamefont {R.}~\bibnamefont {Griessen}},\ }\href@noop {}
  {\bibfield  {journal} {\bibinfo  {journal} {J Alloy Compd}\ }\textbf
  {\bibinfo {volume} {404}},\ \bibinfo {pages} {699} (\bibinfo {year}
  {2005})}\BibitemShut {NoStop}%
\bibitem [{\citenamefont {Dam}\ \emph {et~al.}(2007)\citenamefont {Dam},
  \citenamefont {Gremaud}, \citenamefont {Broedersz},\ and\ \citenamefont
  {Materialia}}]{Dam:2007ip}%
  \BibitemOpen
  \bibfield  {author} {\bibinfo {author} {\bibfnamefont {B.}~\bibnamefont
  {Dam}}, \bibinfo {author} {\bibfnamefont {R.}~\bibnamefont {Gremaud}},
  \bibinfo {author} {\bibfnamefont {C.}~\bibnamefont {Broedersz}}, \ and\
  \bibinfo {author} {\bibfnamefont {R.~G.~S.}\ \bibnamefont {Materialia}},\
  }\href@noop {} {\bibfield  {journal} {\bibinfo  {journal} {Elsevier}\
  }\textbf {\bibinfo {volume} {56}},\ \bibinfo {pages} {853} (\bibinfo {year}
  {2007})}\BibitemShut {NoStop}%
\bibitem [{\citenamefont {Gremaud}\ \emph {et~al.}(2007)\citenamefont
  {Gremaud}, \citenamefont {Slaman}, \citenamefont {Schreuders}, \citenamefont
  {Dam},\ and\ \citenamefont {Griessen}}]{9928980}%
  \BibitemOpen
  \bibfield  {author} {\bibinfo {author} {\bibfnamefont {R.}~\bibnamefont
  {Gremaud}}, \bibinfo {author} {\bibfnamefont {M.}~\bibnamefont {Slaman}},
  \bibinfo {author} {\bibfnamefont {H.}~\bibnamefont {Schreuders}}, \bibinfo
  {author} {\bibfnamefont {B.}~\bibnamefont {Dam}}, \ and\ \bibinfo {author}
  {\bibfnamefont {R.}~\bibnamefont {Griessen}},\ }\href@noop {} {\bibfield
  {journal} {\bibinfo  {journal} {Appl. Phys. Lett.}\ }\textbf {\bibinfo
  {volume} {91}},\ \bibinfo {pages} {231916} (\bibinfo {year}
  {2007})}\BibitemShut {NoStop}%
\bibitem [{\citenamefont {Prinz}\ \emph {et~al.}(2010)\citenamefont {Prinz},
  \citenamefont {P\'alsson}, \citenamefont {Korelis},\ and\ \citenamefont
  {Hj{\"o}rvarsson}}]{prinz:251910}%
  \BibitemOpen
  \bibfield  {author} {\bibinfo {author} {\bibfnamefont {J.}~\bibnamefont
  {Prinz}}, \bibinfo {author} {\bibfnamefont {G.~K.}\ \bibnamefont
  {P\'alsson}}, \bibinfo {author} {\bibfnamefont {P.~T.}\ \bibnamefont
  {Korelis}}, \ and\ \bibinfo {author} {\bibfnamefont {B.}~\bibnamefont
  {Hj{\"o}rvarsson}},\ }\href@noop {} {\bibfield  {journal} {\bibinfo
  {journal} {Appl. Phys. Lett.}\ }\textbf {\bibinfo {volume} {97}},\ \bibinfo
  {pages} {251910} (\bibinfo {year} {2010})}\BibitemShut {NoStop}%
\bibitem [{\citenamefont {Liebig}\ \emph {et~al.}(2008)\citenamefont {Liebig},
  \citenamefont {Andersson}, \citenamefont {Birch},\ and\ \citenamefont
  {Hj{\"o}rvarsson}}]{Liebig:2008bx}%
  \BibitemOpen
  \bibfield  {author} {\bibinfo {author} {\bibfnamefont {A.}~\bibnamefont
  {Liebig}}, \bibinfo {author} {\bibfnamefont {G.}~\bibnamefont {Andersson}},
  \bibinfo {author} {\bibfnamefont {J.}~\bibnamefont {Birch}}, \ and\ \bibinfo
  {author} {\bibfnamefont {B.}~\bibnamefont {Hj{\"o}rvarsson}},\ }\href@noop {}
  {\bibfield  {journal} {\bibinfo  {journal} {Thin Solid Films}\ }\textbf
  {\bibinfo {volume} {516}},\ \bibinfo {pages} {8468} (\bibinfo {year}
  {2008})}\BibitemShut {NoStop}%
\bibitem [{\citenamefont {Droulias}\ \emph {et~al.}(2017)\citenamefont
  {Droulias}, \citenamefont {P\'alsson}, \citenamefont {Palonen}, \citenamefont
  {Hasan}, \citenamefont {Leifer}, \citenamefont {Kapaklis}, \citenamefont
  {Hj\"orvarsson},\ and\ \citenamefont {Wolff}}]{Droulias2017}%
  \BibitemOpen
  \bibfield  {author} {\bibinfo {author} {\bibfnamefont {S.~A.}\ \bibnamefont
  {Droulias}}, \bibinfo {author} {\bibfnamefont {G.~K.}\ \bibnamefont
  {P\'alsson}}, \bibinfo {author} {\bibfnamefont {H.}~\bibnamefont {Palonen}},
  \bibinfo {author} {\bibfnamefont {A.}~\bibnamefont {Hasan}}, \bibinfo
  {author} {\bibfnamefont {K.}~\bibnamefont {Leifer}}, \bibinfo {author}
  {\bibfnamefont {V.}~\bibnamefont {Kapaklis}}, \bibinfo {author}
  {\bibfnamefont {B.}~\bibnamefont {Hj\"orvarsson}}, \ and\ \bibinfo {author}
  {\bibfnamefont {M.}~\bibnamefont {Wolff}},\ }\href {\doibase
  https://doi.org/10.1016/j.tsf.2017.07.005} {\bibfield  {journal} {\bibinfo
  {journal} {Thin Solid Films}\ }\textbf {\bibinfo {volume} {636}},\ \bibinfo
  {pages} {608 } (\bibinfo {year} {2017})}\BibitemShut {NoStop}%
\bibitem [{\citenamefont {Olsson}\ and\ \citenamefont
  {Hj{\"o}rvarsson}(2005)}]{Olsson:2005ii}%
  \BibitemOpen
  \bibfield  {author} {\bibinfo {author} {\bibfnamefont {S.}~\bibnamefont
  {Olsson}}\ and\ \bibinfo {author} {\bibfnamefont {B.}~\bibnamefont
  {Hj{\"o}rvarsson}},\ }\href@noop {} {\bibfield  {journal} {\bibinfo
  {journal} {Phys. Rev. B}\ }\textbf {\bibinfo {volume} {71}},\ \bibinfo
  {pages} {035414} (\bibinfo {year} {2005})}\BibitemShut {NoStop}%
\bibitem [{\citenamefont {Birch}\ \emph {et~al.}(1995)\citenamefont {Birch},
  \citenamefont {Sundgren},\ and\ \citenamefont {Fewster}}]{Birch1995a}%
  \BibitemOpen
  \bibfield  {author} {\bibinfo {author} {\bibfnamefont {J.}~\bibnamefont
  {Birch}}, \bibinfo {author} {\bibfnamefont {J.-E.}\ \bibnamefont {Sundgren}},
  \ and\ \bibinfo {author} {\bibfnamefont {P.}~\bibnamefont {Fewster}},\
  }\href@noop {} {\bibfield  {journal} {\bibinfo  {journal} {J. Appl. Phys.}\
  }\textbf {\bibinfo {volume} {78}},\ \bibinfo {pages} {6562} (\bibinfo {year}
  {1995})}\BibitemShut {NoStop}%
\bibitem [{\citenamefont {Bj{\"{o}}rck}\ and\ \citenamefont
  {Andersson}(2007)}]{Bjorck:aj5091}%
  \BibitemOpen
  \bibfield  {author} {\bibinfo {author} {\bibfnamefont {M.}~\bibnamefont
  {Bj{\"{o}}rck}}\ and\ \bibinfo {author} {\bibfnamefont {G.}~\bibnamefont
  {Andersson}},\ }\href@noop {} {\bibfield  {journal} {\bibinfo  {journal} {J.
  Appl. Crystallogr.}\ }\textbf {\bibinfo {volume} {40}},\ \bibinfo {pages}
  {1174} (\bibinfo {year} {2007})}\BibitemShut {NoStop}%
\bibitem [{\citenamefont {Vorobiev}\ \emph {et~al.}(2015)\citenamefont
  {Vorobiev}, \citenamefont {Devishvilli}, \citenamefont {P\'alsson},
  \citenamefont {Rundlof}, \citenamefont {Johansson}, \citenamefont {Olsson},
  \citenamefont {Dennison}, \citenamefont {Wolff}, \citenamefont {Giroud},
  \citenamefont {Aguettaz},\ and\ \citenamefont {Hjorvarsson}}]{superadam}%
  \BibitemOpen
  \bibfield  {author} {\bibinfo {author} {\bibfnamefont {A.}~\bibnamefont
  {Vorobiev}}, \bibinfo {author} {\bibfnamefont {A.}~\bibnamefont
  {Devishvilli}}, \bibinfo {author} {\bibfnamefont {G.}~\bibnamefont
  {P\'alsson}}, \bibinfo {author} {\bibfnamefont {H.}~\bibnamefont {Rundlof}},
  \bibinfo {author} {\bibfnamefont {N.}~\bibnamefont {Johansson}}, \bibinfo
  {author} {\bibfnamefont {A.}~\bibnamefont {Olsson}}, \bibinfo {author}
  {\bibfnamefont {A.}~\bibnamefont {Dennison}}, \bibinfo {author}
  {\bibfnamefont {M.}~\bibnamefont {Wolff}}, \bibinfo {author} {\bibfnamefont
  {B.}~\bibnamefont {Giroud}}, \bibinfo {author} {\bibfnamefont
  {O.}~\bibnamefont {Aguettaz}}, \ and\ \bibinfo {author} {\bibfnamefont
  {B.}~\bibnamefont {Hjorvarsson}},\ }\href {\doibase
  10.1080/10448632.2015.1057054} {\bibfield  {journal} {\bibinfo  {journal}
  {Neutron News}\ }\textbf {\bibinfo {volume} {26}},\ \bibinfo {pages} {25}
  (\bibinfo {year} {2015})}\BibitemShut {NoStop}%
\bibitem [{elk()}]{elk}%
  \BibitemOpen
  \href@noop {} {\enquote {\bibinfo {title} {The elk fp-lapw code},}\ }\bibinfo
  {note} {Http://elk.sourceforge.net/}\BibitemShut {NoStop}%
\bibitem [{\citenamefont {Romaniello}\ \emph {et~al.}(2006)\citenamefont
  {Romaniello}, \citenamefont {de~Boeij}, \citenamefont {Carbone},\ and\
  \citenamefont {van~der Marel}}]{Romaniello:2006ve}%
  \BibitemOpen
  \bibfield  {author} {\bibinfo {author} {\bibfnamefont {P.}~\bibnamefont
  {Romaniello}}, \bibinfo {author} {\bibfnamefont {P.~L.}\ \bibnamefont
  {de~Boeij}}, \bibinfo {author} {\bibfnamefont {F.}~\bibnamefont {Carbone}}, \
  and\ \bibinfo {author} {\bibfnamefont {D.}~\bibnamefont {van~der Marel}},\
  }\href@noop {} {\bibfield  {journal} {\bibinfo  {journal} {Phys. Rev. B}\
  }\textbf {\bibinfo {volume} {73}},\ \bibinfo {pages} {075115} (\bibinfo
  {year} {2006})}\BibitemShut {NoStop}%
\bibitem [{\citenamefont {Laurent}\ and\ \citenamefont
  {Callaway}(1978)}]{Laurent:1978jq}%
  \BibitemOpen
  \bibfield  {author} {\bibinfo {author} {\bibfnamefont {D.~G.}\ \bibnamefont
  {Laurent}}\ and\ \bibinfo {author} {\bibfnamefont {J.}~\bibnamefont
  {Callaway}},\ }\href@noop {} {\bibfield  {journal} {\bibinfo  {journal}
  {Phys. Rev. B}\ }\textbf {\bibinfo {volume} {17}},\ \bibinfo {pages} {455}
  (\bibinfo {year} {1978})}\BibitemShut {NoStop}%
\bibitem [{\citenamefont {Weaver}\ \emph {et~al.}(1974)\citenamefont {Weaver},
  \citenamefont {Lynch},\ and\ \citenamefont {Olson}}]{Weaver:1974vx}%
  \BibitemOpen
  \bibfield  {author} {\bibinfo {author} {\bibfnamefont {J.}~\bibnamefont
  {Weaver}}, \bibinfo {author} {\bibfnamefont {D.}~\bibnamefont {Lynch}}, \
  and\ \bibinfo {author} {\bibfnamefont {C.}~\bibnamefont {Olson}},\
  }\href@noop {} {\bibfield  {journal} {\bibinfo  {journal} {Phys. Rev. B}\
  }\textbf {\bibinfo {volume} {10}},\ \bibinfo {pages} {501} (\bibinfo {year}
  {1974})}\BibitemShut {NoStop}%
\bibitem [{\citenamefont {Sacchetti}(1983)}]{Sacchetti:1983jl}%
  \BibitemOpen
  \bibfield  {author} {\bibinfo {author} {\bibfnamefont {F.}~\bibnamefont
  {Sacchetti}},\ }\href@noop {} {\bibfield  {journal} {\bibinfo  {journal} {J.
  Phys. F}\ }\textbf {\bibinfo {volume} {13}},\ \bibinfo {pages} {1801}
  (\bibinfo {year} {1983})}\BibitemShut {NoStop}%
\bibitem [{\citenamefont {Palik}(1998)}]{Palik:1998}%
  \BibitemOpen
  \bibfield  {author} {\bibinfo {author} {\bibfnamefont {E.}~\bibnamefont
  {Palik}},\ }\href@noop {} {\emph {\bibinfo {title} {{Handbook of Optical
  Constants of Solids}}}}\ (\bibinfo  {publisher} {Academic Press},\ \bibinfo
  {year} {1998})\BibitemShut {NoStop}%
\bibitem [{\citenamefont {Smithson}\ \emph {et~al.}(2002)\citenamefont
  {Smithson}, \citenamefont {Marianetti}, \citenamefont {Morgan}, \citenamefont
  {Van~der Ven}, \citenamefont {Predith},\ and\ \citenamefont
  {Ceder}}]{Smithson:2002tla}%
  \BibitemOpen
  \bibfield  {author} {\bibinfo {author} {\bibfnamefont {H.}~\bibnamefont
  {Smithson}}, \bibinfo {author} {\bibfnamefont {C.~A.}\ \bibnamefont
  {Marianetti}}, \bibinfo {author} {\bibfnamefont {D.}~\bibnamefont {Morgan}},
  \bibinfo {author} {\bibfnamefont {A.}~\bibnamefont {Van~der Ven}}, \bibinfo
  {author} {\bibfnamefont {A.}~\bibnamefont {Predith}}, \ and\ \bibinfo
  {author} {\bibfnamefont {G.}~\bibnamefont {Ceder}},\ }\href@noop {}
  {\bibfield  {journal} {\bibinfo  {journal} {Phys. Rev. B.}\ }\textbf
  {\bibinfo {volume} {66}},\ \bibinfo {pages} {144107} (\bibinfo {year}
  {2002})}\BibitemShut {NoStop}%
\bibitem [{\citenamefont {P\'alsson}\ \emph {et~al.}(2014)\citenamefont
  {P\'alsson}, \citenamefont {Eriksson}, \citenamefont {Amft}, \citenamefont
  {Xin}, \citenamefont {Liebig}, \citenamefont {\'Olafsson},\ and\
  \citenamefont {Hj\"orvarsson}}]{PhysRevB.90.045420}%
  \BibitemOpen
  \bibfield  {author} {\bibinfo {author} {\bibfnamefont {G.~K.}\ \bibnamefont
  {P\'alsson}}, \bibinfo {author} {\bibfnamefont {A.~K.}\ \bibnamefont
  {Eriksson}}, \bibinfo {author} {\bibfnamefont {M.}~\bibnamefont {Amft}},
  \bibinfo {author} {\bibfnamefont {X.}~\bibnamefont {Xin}}, \bibinfo {author}
  {\bibfnamefont {A.}~\bibnamefont {Liebig}}, \bibinfo {author} {\bibfnamefont
  {S.}~\bibnamefont {\'Olafsson}}, \ and\ \bibinfo {author} {\bibfnamefont
  {B.}~\bibnamefont {Hj\"orvarsson}},\ }\href@noop {} {\bibfield  {journal}
  {\bibinfo  {journal} {Phys. Rev. B}\ }\textbf {\bibinfo {volume} {90}},\
  \bibinfo {pages} {045420} (\bibinfo {year} {2014})}\BibitemShut {NoStop}%
\end{thebibliography}%

\end{document}